\documentclass[pdftex,onecolumn]{pasj01}
\Received{$\langle$reception date$\rangle$}
\Accepted{$\langle$acception date$\rangle$}
\Published{$\langle$publication date$\rangle$}
\begin{document}

\title{First clear detection of the CCS Zeeman splitting toward the pre-stellar core, Taurus Molecular Cloud-1}
\author{Fumitaka \textsc{Nakamura}\altaffilmark{1,2,3},
Seiji \textsc{Kameno}\altaffilmark{1,4},
Takayoshi \textsc{Kusune}\altaffilmark{1},
Izumi \textsc{Mizuno}\altaffilmark{5},
Kazuhito \textsc{Dobashi}\altaffilmark{6},
Tomomi \textsc{Shimoikura}\altaffilmark{6},
Kotomi \textsc{Taniguchi}\altaffilmark{7}}

\altaffiltext{1}{National Astronomical Observatory of Japan, 2-21-1 Osawa, Mitaka, Tokyo 181-8588, Japan}
\altaffiltext{2}{The Graduate University for Advanced Studies (SOKENDAI), 2-21-1 Osawa, Mitaka, Tokyo 181-0015, Japan}
\altaffiltext{3}{The University of Tokyo, 7-3-1 Hongo Bunkyo, 113-0033 Tokyo, Japan}
\altaffiltext{4}{Joint ALMA Observatory, Alonso de C\'ordova 3107 Vitacura, Santiago, Chile}
\altaffiltext{5}{East Asian Observatory, 660 N. Aohoku Place, University Park, Hilo, Hawaii 96720, U.S.A.
}
\altaffiltext{6}{Department of Astronomy and Earth Sciences, Tokyo Gakugei University, 4-1-1 Nukuikitamachi, Koganei, Tokyo 184-8501, Japan}
\altaffiltext{7}{Department of Astronomy, University of Virginia, P.O. Box 3818, Charlottesville, VA 22903-0818}
\email{fumitaka.nakamura@nao.ac.jp}


\maketitle

\begin{abstract}
We report a first clear detection of the Zeeman splitting of a CCS emission line at 45 GHz toward 
a nearby prestellar dense filament, Taurus Molecular Cloud-1.  We observed HC$_3$N 
non-Zeeman line simultaneously as the CCS line, and did not detect any significant splitting of HC$_3$N line.
Thus, we conclude that our detection of the CCS Zeeman splitting is robust.
The derived \textcolor{black}{line-of-sight} magnetic field strength is about 117 $\pm$ 21 $\mu$G, 
which corresponds to the normalized mass-to-magnetic flux ratio of 2.2 if we adopt the 
inclination angle of 45$^\circ$. Thus, we conclude that the TMC-1 filament is magnetically supercritical.
Recent radiative transfer calculations of CCS and HC$_3$N lines along the line of sight suggest 
that the filament is collapsing with a speed of $\sim$ 0.6 km s$^{-1}$, 
which is comparable to three times the isothermal sound speed.
This infall velocity appears to be consistent with the evolution of a gravitationally-infalling core.
\end{abstract}

\section{Introduction}
\label{sec:intro}

Stars form by gravitational contraction of dense parts of molecular clouds called
pre-stellar cores \citep{shu87,benson89,bergin07}.
During their gravitational contraction, the magnetic field influences
their dynamical evolution through extracting angular momentum by 
magnetic braking, driving protostellar jets, and
suppressing fragmentation, 
and thus controls the formation and evolution of protostars and protoplanetary discs \citep{li14,zhao16}. 
The magnetic field also influences the timescale and efficiency of star 
and planet formation by impeding the contraction \citep{li04}.
Therefore, measuring the magnetic field strength of pre-stellar cores
is crucial to understand how stars and planets are formed in our Galaxy.
However, the measurements of the magnetic strength of pre-stellar cores 
have been very limited because of technical difficulties \citep{crutcher12}.
%%%% 
Thus, there is debate regarding the dynamical importance 
of magnetic field on core evolution \citep{tasis09,li14b}.

%%%%%
%Here we report the detection of 117 $\pm$ 21 $\mu$ Gauss magnetic field 
%toward a pre-stellar core, TMC-1, through CCS (thioxoethenylidene) Zeeman 
%measurements using a newly-developed receiver system installed on the Nobeyama 45-m telescope\citep{mizuno14,nakamura14}.
%This is the first clear detection of the Zeeman splitting of the CCS emission line in interstellar space.
%The derived field strength indicates that TMC-1 is magnetically supercritical.
%Thus, the magnetic field cannot prevent completely from global gravitational contraction, 
%suggesting that TMC-1 is on the verge of protostellar formation. 
%TMC-1 is magnetically-supported 
%against gravity.
%%%% 
%The existence of such a core supports a scenario of  magnetically-regulated star formation\cite{li14b}, 
%and may resolve a long standing problem of star formation,
%%%
%The existence of such a core contradicts the current picture of
%star formation\cite{crutcher12}, but may resolve a long standing problem of %star formation, 
%how to remove angular momentum from protostars. 

Most reliable magnetic field measurements toward molecular clouds 
are done by the Zeeman observations using molecular lines \citep{crutcher12}.
The degenerate energy state of a molecule with an unpaired electron is resolved under magnetic fields and yields a line splitting between orthogonal circular polarizations.
Until now, the Zeeman measurements towards star-forming regions 
are mainly carried out with H{\scriptsize I}, OH, and CN and some maser lines \citep{fiebig89,crutcher12}.  
There are, however, several disadvantages of these measurements.
The H{\scriptsize I} and OH lines trace only low-density envelopes of $10^2-10^3$ cm$^{-3}$ because of
their low critical densities. 
In addition, the spatial resolution is poor due to their low frequencies of $\sim 1$ GHz
even with the largest telescopes such as the Arecibo. 
Although CN lines can trace a higher density of $\sim 10^5$ cm$^{-3}$, 
the CN Zeeman observations have been limited primarily to {\it protostellar} 
regions with strong CN emission. The CH$_3$OH and OH maser lines come from small
compact spots close to protostars, and may not trace the core magnetic fields.
Thus, the measurements of the magnetic strengths of dense molecular cloud cores, particularly cores prior to star formation 
or pre-stellar cores in short, have been extremely limited.

The typical density of pre-stellar cores is about $10^{4-5}$ cm$^{-3}$.
At this density range, there are only a few appropriate Zeeman lines.
One of the appropriate lines is the rotational transition of CCS, whose
critical density is around $10^4$ cm$^{-3}$.
To directly measure the field strength of pre-stellar cores, 
we developed a new 45-GHz dual-linear polarization receiver installed on the Nobeyama 45-m telescope called Z45 \citep{nakamura15,mizuno14}, 
and carried out the Zeeman observations of CCS ($J_N=4_3-3_2$,  45.379033 GHz, Yamamoto et al. 1990). 
The dicarbon monosulfide (CCS) is a radical molecule, having an unpaired electron
and therefore the Lande factor is relatively large \citep{shinnaga00}.
In addition, CCS is abundant in pre-stellar phase \citep{suzuki92,marka12,nakamura14,shimoikura18,shimoikura19}
and thus it is suitable for the Zeeman measurements towards pre-stellar cores.
However, previous Zeeman observations using CCS \citep{shinnaga99,levin01}
do not give successful detection of the Zeeman splitting in a satisfactory level
due to both the insufficient signal-to-noise ratio ($\sim 1$) and technical difficulties 
in the removal of the instrumental polarization contribution.  
Note that \citet{turner06} attempted to perform the Zeeman measurements toward TMC-1 
with C$_4$H, and reported the upper limit of 14.5 $\pm$ 14 $\mu$G.
We overcame these two problems by developing a new observation system 
\citep{mizuno14, nakamura15}. 
In the present paper we report a first most reliable detection of the Zeeman splitting of the 
CCS ($J_N=4_3-3_2$) line
toward a prestellar core, Taurus Molecular Cloud 1 (hereafter, TMC-1).

The paper is organized as follows. Section \ref{sec:obs} describes the details of the observations.
In section \ref{sec:results}, we present our Zeeman measurements toward TMC-1. In section 
\ref{sec:discussion}
we discuss how the magnetic field plays an important role in the dynamics of the TMC-1. Finally, we summarize 
our main results in section \ref{sec:summary}.

\section{Observations}
\label{sec:obs}

\subsection{Position-switch polarization observations}

Observations were carried out during the period of  April 2014 $-$ May 2015 
in position-switch mode using the Z45 receiver installed in the Nobeyama 45-m telescope \citep{nakamura15}.
The observed core is TMC-1, a pre-stellar filamentary core.
The mean distance to the Taurus association is estimated to be 137 pc 
based on the VLBI observations \citep{torres07}. 
Hereafter, we adopt the distance of 140 pc.
We chose the position of the  CCS peak intensity at the integrated intensity map, and 
the coordinate of the target position is 
[R.A.(J2000.0), Decl.(J2000.0)]= (4$^h$ 41$^m$ 43$^s$.87, 25$^\circ$ 41\arcmin 17\arcsec.7).  
We chose a position $+$30$'$ offset in the R.A. direction as an emission free position.

\begin{figure}
\begin{centering}
\includegraphics[width=1.0\textwidth]{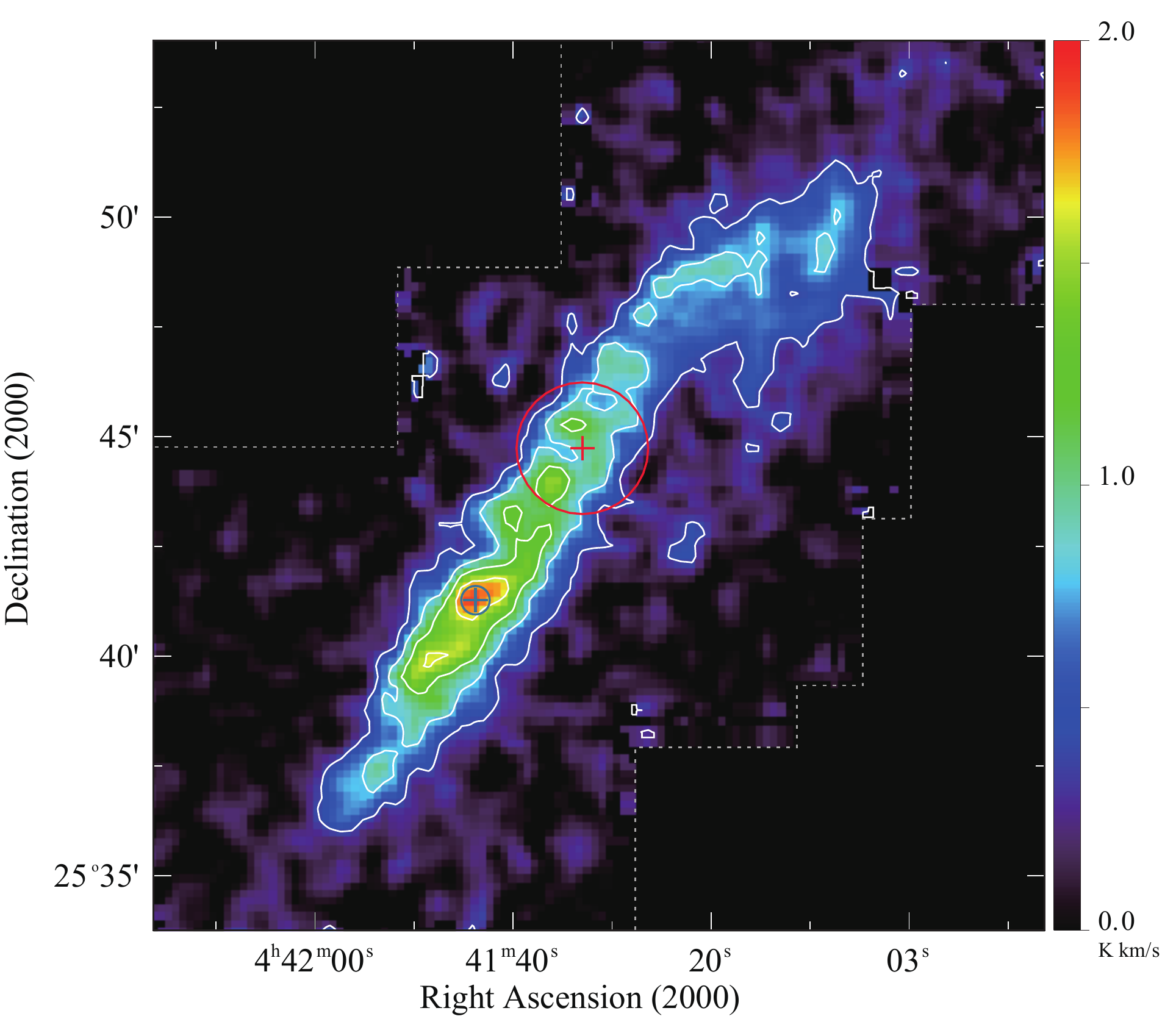}
\end{centering}
\caption{Total integrated intensity map of CCS ($J_N=4_3-3_2$) toward TMC-1.
Our target position and the position of OH Zeeman observations \citep{troland08} 
are shown in the black and red plus signs, respectively.
The sizes of the circles indicate the HPBW beam sizes.}
\label{fig:tmc-1}
\end{figure}

TMC-1 has one of the strongest CCS emission in the Taurus molecular cloud
\citep{suzuki92}.
In figure \ref{fig:tmc-1} we show the CCS integrated intensity map of TMC-1.
The observed position marked with a black circle in figure \ref{fig:tmc-1}  is 
located in the western part of the TMC-1 filament.
We observed two molecular  lines of CCS ($J_N=4_3-3_2$) 
and HC$_3$N ($J=5-4$, 45.490316 GHz, Lafferty \& Lovas 1978) simultaneously.
The HC$_3$N line is a non-Zeeman one, and is used 
\textcolor{black}{to check how well the polarization calibration is performed}. 
In addition, the HC$_3$N line traces the same density range as the CCS line \citep{suzuki92,taniguchi18}.
As a backend, we used a newly-developed software spectrometer, PolariS \citep{mizuno14}, 
which has a frequency resolution of 54 Hz (FWHM).
The beam size (HPBW) of the Z45 receiver is 40\arcsec at 45 GHz, which corresponds to 0.024 pc
\textcolor{black}{at the location of TMC-1}. 
The typical noise temperature was in the range of  100 $-$ 200 K. 
The pointing was checked every 1.5 hour with the SiO maser of NML Tau.

%To reduce the total observation time, we applied the smoothed bandpass calibration (SBC)\citep{yamaki12}.  
%In the SBC method, we reduce the integration time
%of the emission-free position and smoothed the data of the position with a third-order B-spline smoothing 
%to reduce random noise at the emission-free position.  In our observations, 
%SBC allowed us to reduce the total observation time by a factor of three.

We adopted smoothed bandpass calibration (SBC) method \citep{yamaki12} with which we can reduce
the integration time of \textcolor{black}{off-source scans by smoothing the spectrum at the off-source blank sky position (the reference position). Since the appropriate spectrum smoothing reduces the noise, we can shorten the integration time for the reference position.}  The integration time was 120 s and 10 s for the target position and \textcolor{black}{reference position}, respectively. 
We smoothed \textcolor{black}{the spectrum at the reference position using a third-order spline function 
as the baseline for obtaining the spectrum of molecular lines at the target position}.
In our Zeeman observations, the SBC method allowed us to reduce the total observation time 
by a factor of three.

The details of the calibration, beam squint correction, and data analysis will be shown in a forthcoming paper, and
some data analysis procedures of the polarization data are summarized in the appendix.

\subsection{OTF observations}

We also carried out mapping observations of TMC-1 in the on-the-fly (OTF) mode with the Z45 receiver and a digital
spectrometer, SAM45 \citep{kamazaki12}. The frequency resolution was set to 3.81 kHz, 
corresponding to the velocity resolution of 0.025 km s$^{-1}$.
The typical noise temperature was around 100 $-$ 200 K. The pointing was checked every 1.5 hour 
\textcolor{black}{by observing} the SiO maser of NML Tau.
The noise level of the map is estimated to be about 0.075 K in a brightness temperature scale using the main beam efficiency of 0.7.
CCS ($J_N=4_3-3_2$) and HC$_3$N ($J=5-4$) lines are simultaneously observed.
The map is used to measure the velocity gradient at the observed position.
The details of the observations are also described in \citet{dobashi19}.

\section{Results}
\label{sec:results}

\begin{figure}
\begin{centering}
\includegraphics[width=80mm, bb = 0 0 800 800]{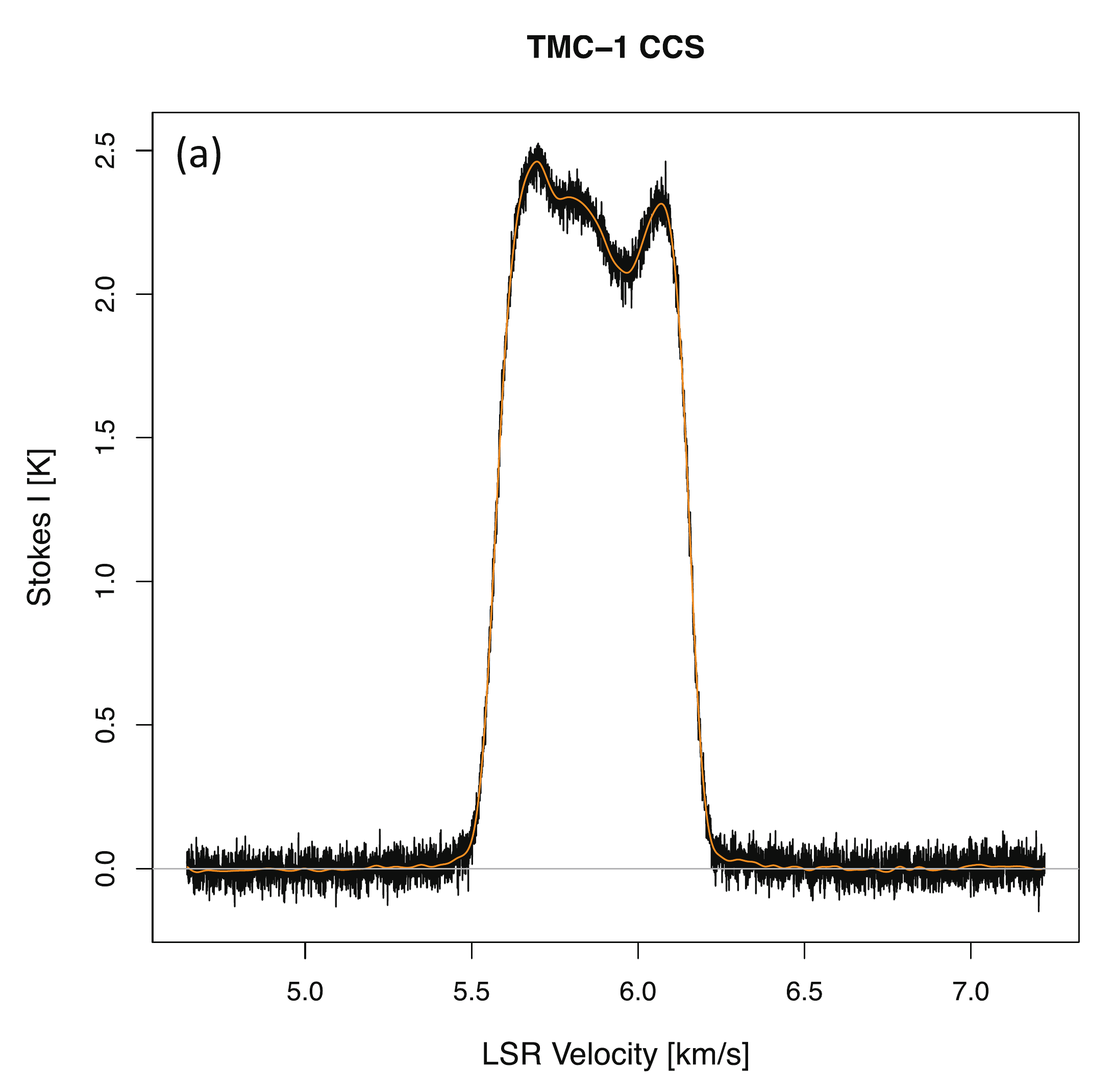}
\includegraphics[width=80mm, bb = 0 0 800 800]{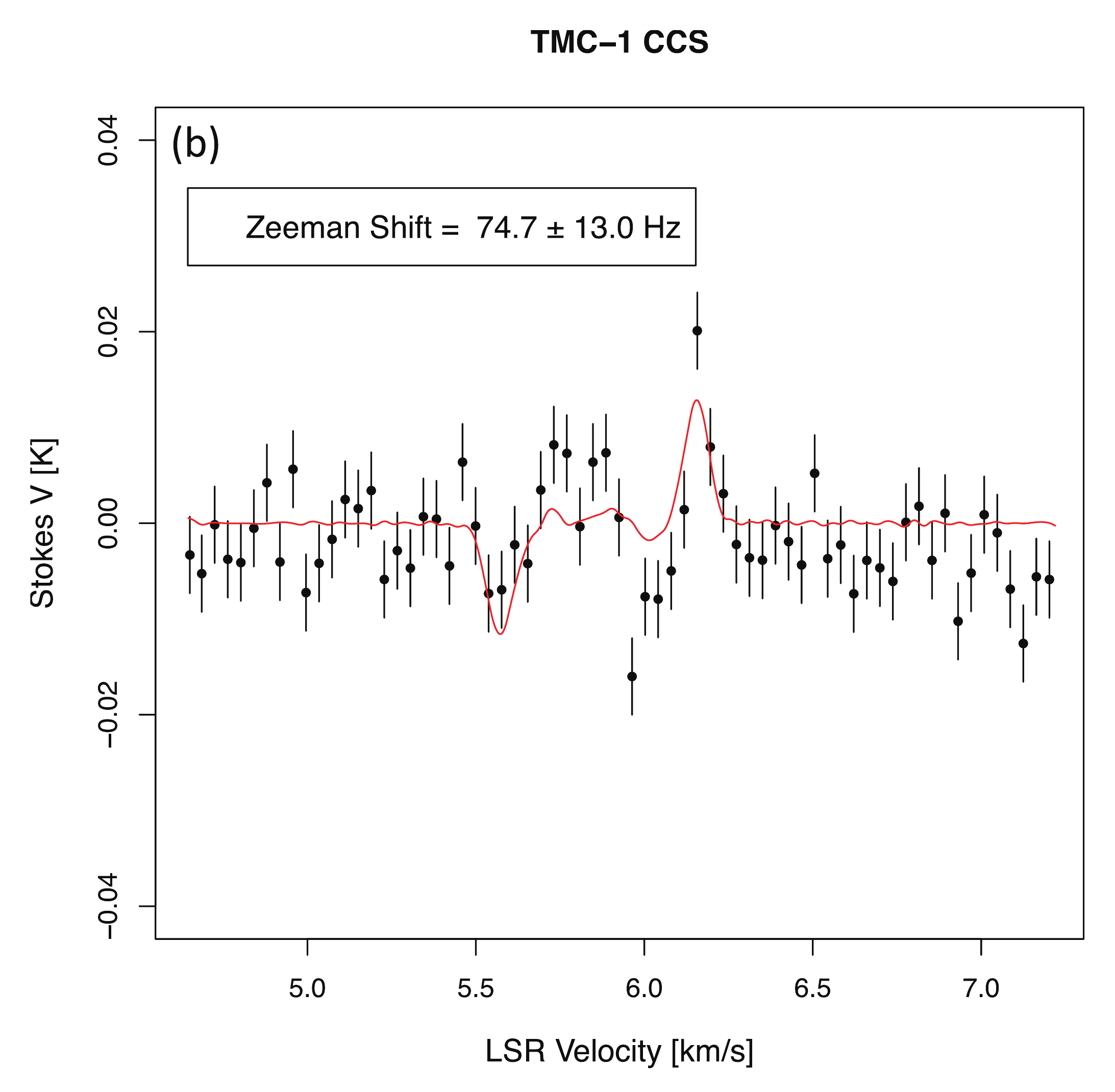}
\end{centering}
\caption{(a) Stokes $I$ profile of CCS toward TMC-1. (b) Stokes $V$ profile of CCS.
The red line shows the fitted Stokes $V$ profile.
The $t$ and $p$ values of the fitting are $t=9.0$ and $p < 2\times 10^{-16}$, respectively.}
\label{fig:ccs}
\end{figure}

\begin{figure}
\includegraphics[width=80mm, bb = 0 0 800 800]{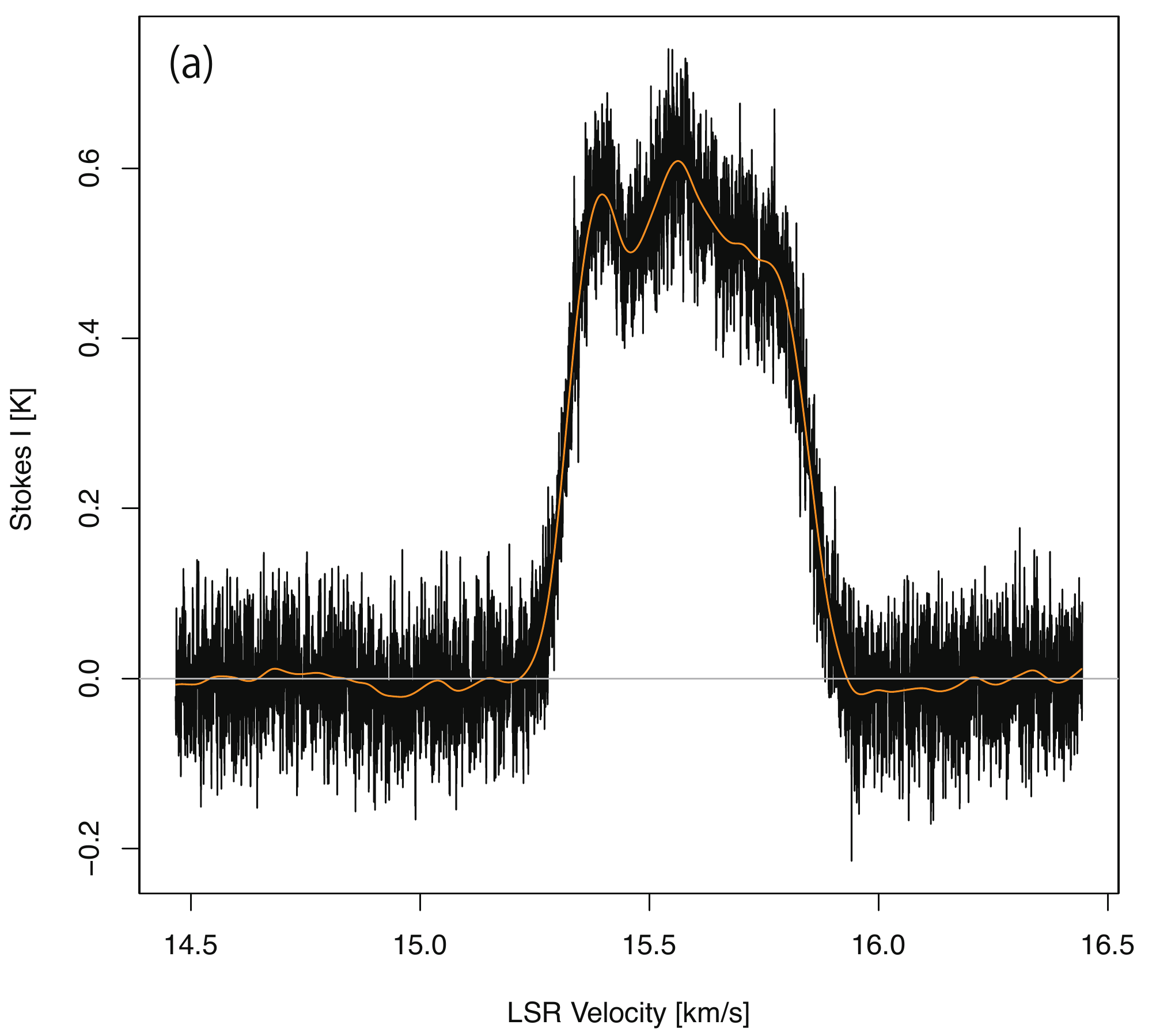}
\includegraphics[width=80mm, bb = 0 0 800 800]{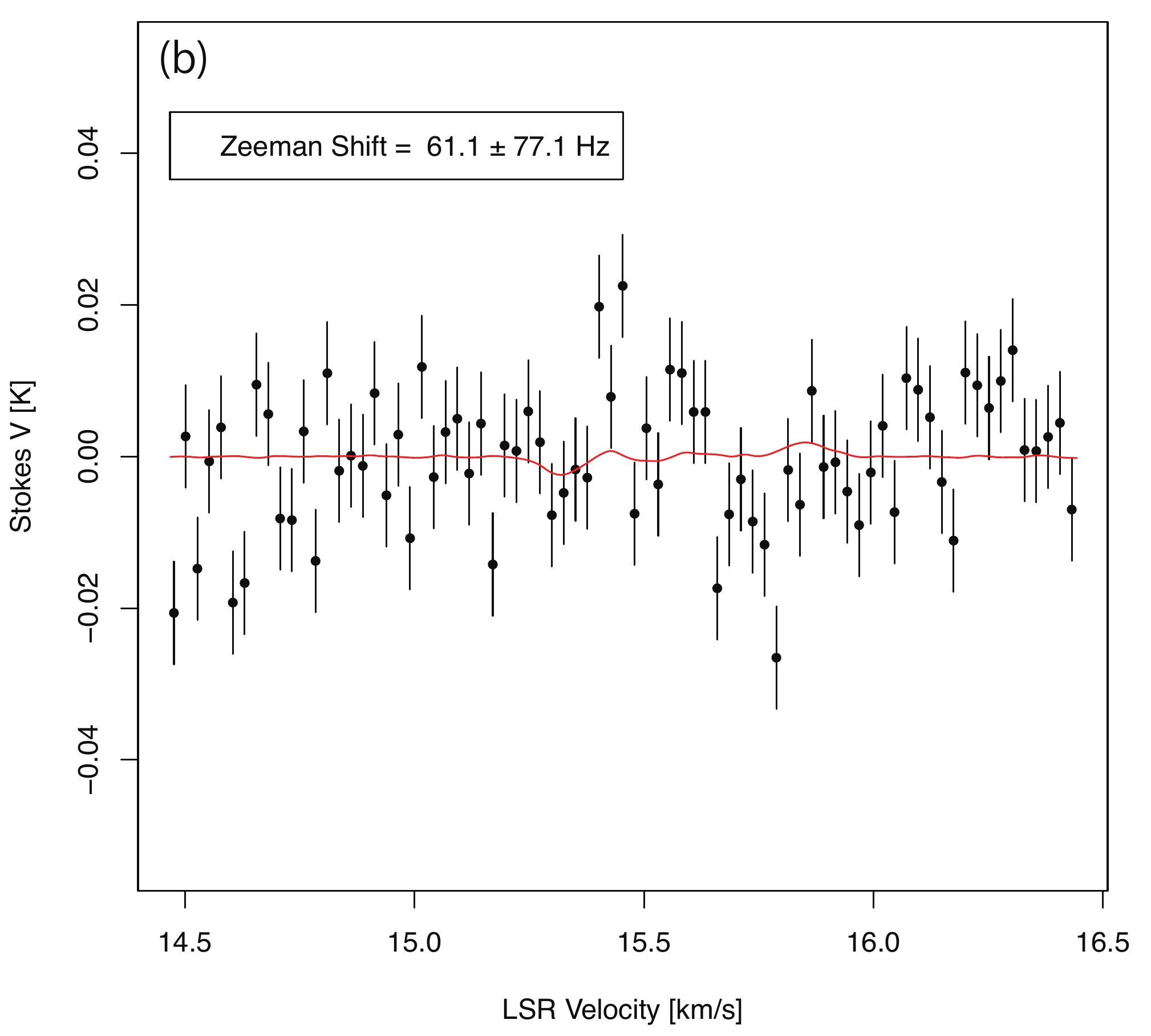}
\caption{(a) Stokes $I$ profile of the HC$_3$N  ($J=5-4$) satellite component ($F=4-4$) toward TMC-1.
Note that the satellite component does not have blended hyperfine components and optically thin.
The rest frequency of the HC$_3$N ($J=5-4, F=5-4$) line was adopted to measure the line-of-sight velocity.
%The three hyperfine components are blended for the main component.
(b) Stokes $V$ profile of HC$_3$N. The red line shows the fitted Stokes V profile.
The $t$ and $p$ values of the fitting were $t=0.7$ and $p=0.48$, respectively.
%(c) Same as panel (a) but for the satellite component ($F=4-4)$.  
%(d) Same as panel (b) but for the satellite component ($F=4-4)$.
%Note that the satellite components do not have blended hyperfine components.
%The $t$ and $p$ values of the fitting were $t=0.7$ and $p=0.48$, respectively.
}
\label{fig:hc3n}
\end{figure}

Figures \ref{fig:ccs} and \ref{fig:hc3n} show the Stokes $I$ and $V$ spectra of the CCS and HC$_3$N 
lines toward our observed point, respectively.
%Since our system has a beam squint of 2$"$ between the two linearly polarized components,
We applied the beam squint correction to obtain the Stokes $V$ spectra.
Our system has a beam squint of the two circularly-polarized components of 
about 2$''$ mainly in the azimuth direction, with a dependence on the observed elevation.  
To calculate the splitting, we first fit the observed Stokes $I$ profile with a third-order B-spline function 
and derive $dI/d\nu$ which is proportional to the Stokes $V$ 
caused by the Zeeman split as $V\propto dI/d \nu$ \citep{crutcher93}. 
Then, we carried out a least-square fit to the Stokes $V$ spectra 
with a function of 
\begin{equation}
{\rm Stokes} \ V = a_1+ a_2 I + a_3 {dI \over d\nu}  \ ,
\end{equation}
where the coefficients $a_1$ and $a_2$ are 
almost zero ($a_1 \simeq 6.7 \times 10^{-3}$ K and $a_2 \simeq 1.9\times 10^{-3}$).
The frequency shift of the Stokes $V$, $a_3$, is proportional to the magnetic strength along the line-of-sight direction 
($a_3 \propto B_{\rm los}$), and  we derived the frequency shift of $+75.3$ $\pm$ 13.4 Hz, 
where the second number is the standard error.  
The $t$ and $p$ values were $t=9.0$ and $p < 2\times 10^{-16}$, respectively, suggesting that
$a_3$ is significantly required to explain the observed Stokes $V$ profile. 
From this splitting frequency, we derive the line-of-sight magnetic field strength 
of 117 $\pm$ 21  $\mu$G with the Lande factor of CCS ($J_N=4_3-3_2)$ 
64 Hz /$100 \mu$G \citep{shinnaga00}. 
It is worth noting that the CCS line profile can be fitted with 4 Gaussian components 
with different line-of-sight velocities ($\displaystyle I=\sum _{i=1}^4 I_i$) \citep{dobashi18}. 
Therefore, we implicitly assume that the strength of the magnetic field associated 
with each component is the same,
so that 
\begin{equation}
\displaystyle V=\sum  _{i=1}^4 a_3 {dI_i \over d\nu} = a_3 {dI \over d\nu} \ .
\end{equation}

We applied the same procedure as CCS to the HC$_3$N line simultaneously obtained
and derived a frequency shift of 
%$8.2 \pm 13.5$ Hz and
$61.1 \pm 77.1$ Hz for the %main component and 
satellite component ($F=4-4$).  Here, we did not use the main component since it consists of the three optically thick hyperfine components ($F=5-3$, $5-4$, and $6-5$) which make the total profile very complicated.
The $t$ and $p$ values of the fitting are %$t=0.7$ and $p =0.48$ and 
$t=0.7$ and $p =0.48$.
%, respectively. 
%The Stokes V profile of HC$_3$N appears to have some pattern at the envelope of the profile 
%but it is not due to the Zeeman splitting.
Thus, the derived frequency shift of HC$_3$N is statistically insignificant, and
we conclude that we detect the CCS Zeeman shift but not for HC$_3$N. 
This is the first clear detection of the CCS Zeeman splitting from a prestellar core.

There has been no clear detection reported in previous observations mainly because 
of the technical difficulties such as low signal-to-noise ratios and 
instrumental polarization effects.
There are at least three advantages in our measurements. 
Firstly, we used a {\it linear} polarization receiver 
and took a cross-correlation between linearly-polarized components \citep{heiles01}, 
whereas the previous attempts were done with {\it circular} polarization reception
 that can generate significant systematic errors in Stokes $V$ spectra caused by gain inhomogeneity 
between two-circularly polarized components. 
Secondly, we applied the SBC method which
can improve the signal-to-noise ratio by a factor of about 3 for narrow emission lines within a limited observation time.
Lastly, simultaneous observations of the Zeeman line (CCS) and non-Zeeman line (HC$_3$N)
guarantee our detection.  In addition, we have done simultaneous observations of CCS and another non-Zeeman line (HC$_5$N)
and obtained the similar splitting of the Stokes $V$ for CCS and no significant splitting for HC$_5$N. 
We note that to verify our polarization system, we also observed OMC-2 in  $7_0 - 6_1 A^+$ 
methanol maser line at 44.1 GHz and detected a Zeeman splitting, which was consistent with the previous detection with VLA \citep{sarma11}.

\section{Importance of magnetic field in TMC-1}
\label{sec:discussion}

How significantly does this magnetic field influence the dynamics of the TMC-1 core?
From the Stokes $V$ profiles, we can only derive the line-of-sight strength of the magnetic field.
To obtain the true field strength, we need to estimate the field strength on the plane-of-sky.
It is, however, difficult to accurately derive the strength of  the plane-of-sky component observationally,
although we can roughly evaluate the {\it spatially-averaged} strength of the plane-of-sky component 
from linear polarization observations \citep{crutcher12}.
From near-infrared polarization observations, the average strength of the 
plane-of-sky component toward the TMC-1 region is estimated to be $\left< B_{\rm pos}\right> \sim 61 \ \mu$ G 
at the average density of 970 cm$^{-3}$ 
by applying the modified Chandrasekhar-Fermi method \citep{chapman11}. 
The volume density of H$_2$ at our position is derived 
to be $3 \times 10^4 $ cm$^{-3}$ 
by applying the LVG method using the CCS ($J_N=2_1-1_0$) and ($J_N=4_3-3_2$) 
lines \citep{suzuki92}.
If we scale the above value to the density traced by CCS assuming $B \propto n^{0.5}$,
the expected strength is stronger than the line-of-sight component as 339 $\mu$G.
We consider that this value of $\left<B_{\rm pos}\right>$ may be significantly 
overestimated because the TMC-1 core contains multiple components with different velocities
and may undergo global contraction which increases the local velocity dispersion.

\begin{figure}
\begin{centering}
\includegraphics[width=80mm, bb = 0 0 800 800]{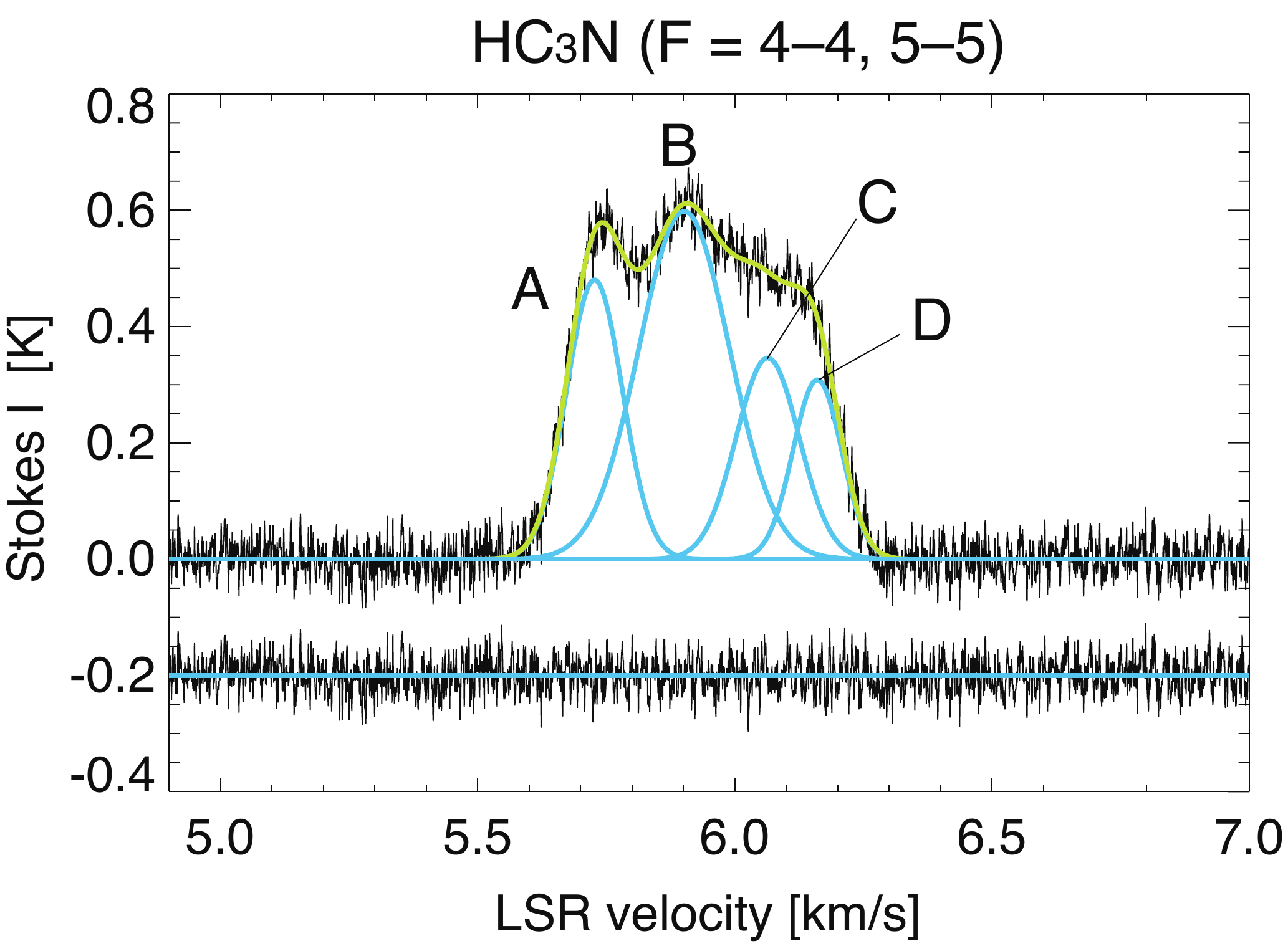} 
%\clearpage
\includegraphics[width=80mm, bb = 0 0 800 800]{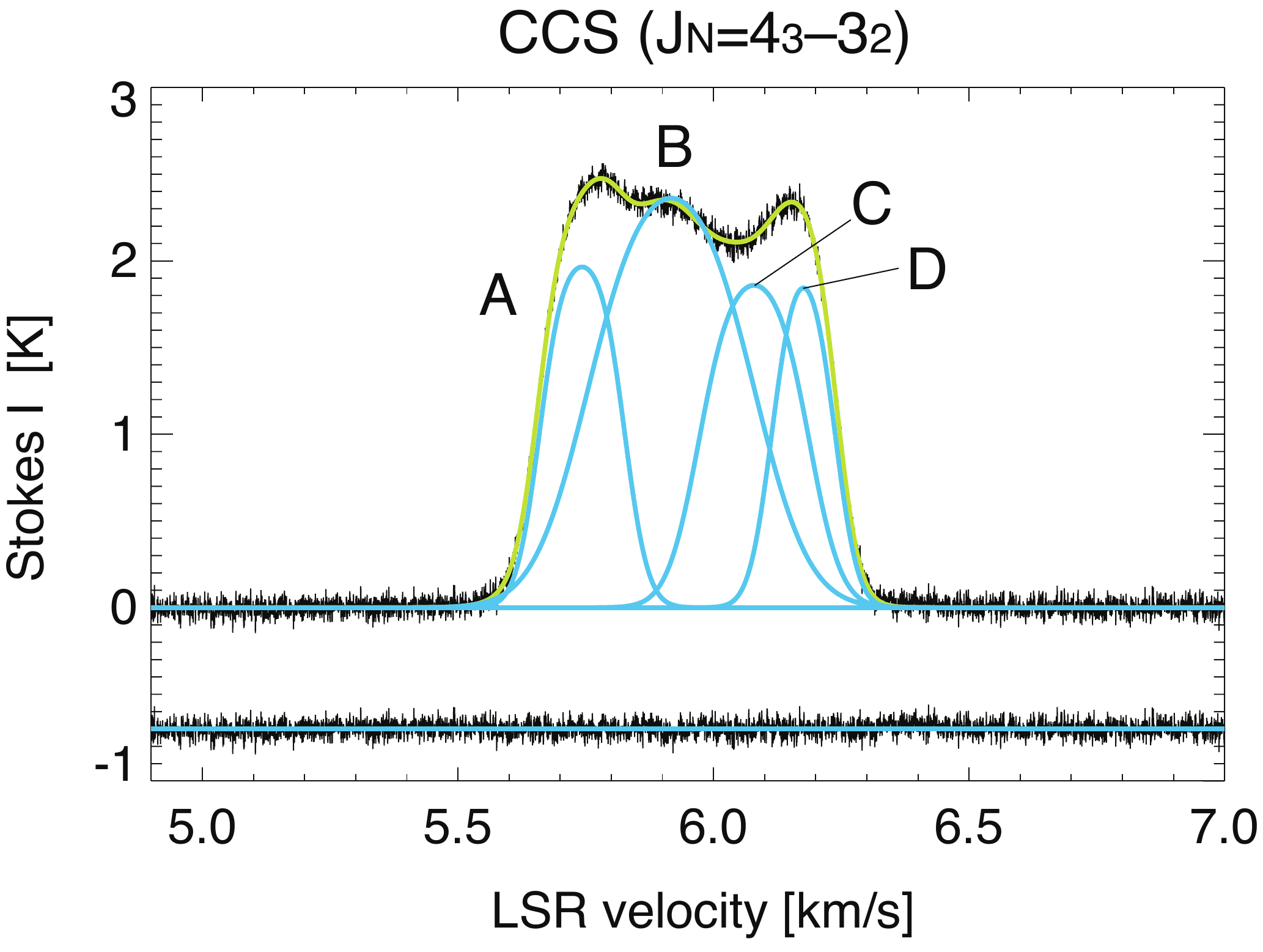}
\end{centering}
\caption{({\it left}) A hyperfine component of the HC$_3$N ($J=5-4$) line fitted with four Gaussian profiles
 (blue lines). The yellow line denotes their simple sum.
The $F=5-5$ and $F=4-4$ profiles are combined to improve the noise level.
The residual spectrum of the Gaussian fitting is shown just below the actual spectrum. 
({\it right}) CCS line profile fitted with four components. The blue lines are the simple Gaussian functions, 
and the yellow line represents their sum calculated taking into account the radiative transfer 
(see Dobashi et al. 2018).
}
\label{fig:fitted}
\end{figure}

\begin{figure}
\includegraphics[width=160mm, bb = 0 0 600 400]{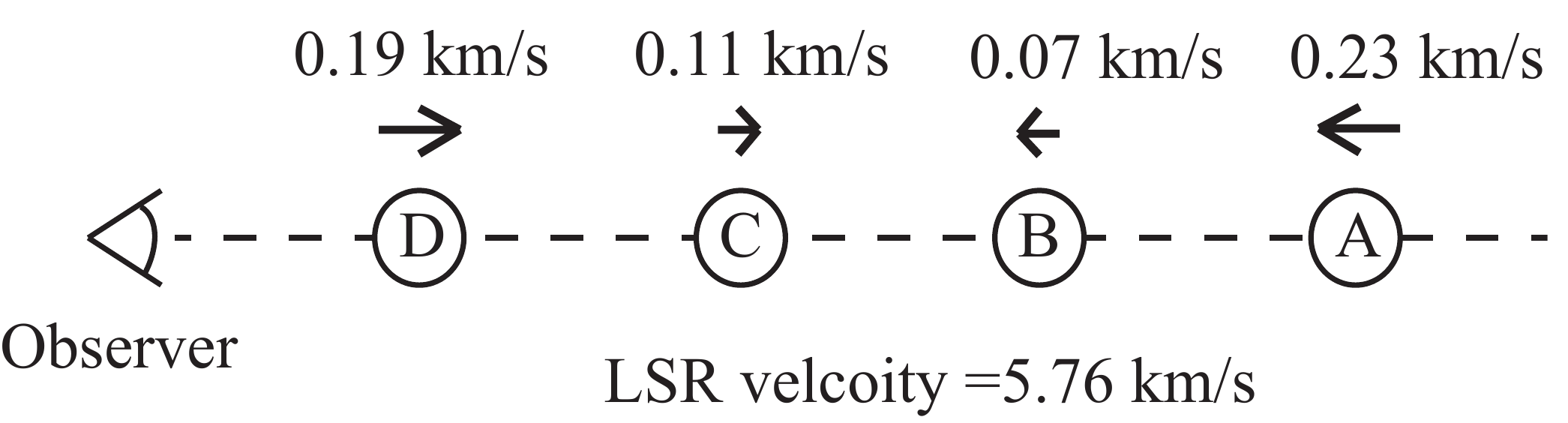}
\caption{Spatial configuration of the four CCS components.  
The velocities indicated by the arrows are the relative velocities with respect to the mean velocity of 5.76 km s$^{-1}$.}
\label{fig:components}
\end{figure}

As mentioned above, at our observed position, the CCS line profile can be fitted with four Gaussian components with different velocities
\citep{dobashi18}.
The four components A, B, C, and D have the centroid velocities of 5.53 km s$^{-1}$, 
5.69 km s$^{-1}$, 5.87 km s$^{-1}$, and 5.95 km s$^{-1}$, respectively, which are determined by Gaussian fitting
of the optically-thin HC$_3$N isolated hyperfine line (see figure \ref{fig:fitted}(a)). 
Based on the detailed radiative transfer calculations, the most plausible spatial configurations of these four components are inferred to 
the one that lines up in order as A, B, C, and D from the furthest position along the line-of-sight 
(see figure \ref{fig:components}), where the components A, B, C, and D are indicated in figure \ref{fig:fitted}(b). 
The velocity difference of the components A and D is about 0.4 km s$^{-1}$, and 
this may be due to the global infall or converging motion that contributes dominantly to the total 
line width of the CCS profile ($\approx 0.58 $km s$^{-1}$).
The existence of such a global converging flow is consistent with large-scale velocity gradient seen in wide-field CO maps \citep{narayanan08}. 
Such a global motion is likely to significantly contribute to the total velocity 
width in this core. Thus, the magnetic strength estimated by the Chandrasekhar-Fermi 
analysis may be overestimated. 
The individual components have the velocity widths of about 0.2 km s$^{-1}$, about a third of the total velocity width.
Therefore, the strength of the plane-of-sky component may be much weaker as 20 $\mu$G at 970 cm$^{-3}$.
Assuming that the field strength increases in proportion to the square root of the \textcolor{black}{number density}, 
$\left<B_{\rm pos}\right>$ of TMC-1 at $n \simeq 3\times 10^4$ cm$^{-3}$ is scaled to be 
$\left<B_{\rm pos}\right> \simeq 100 \  \mu$G. 
The total magnetic field strength at our observed position can be evaluated to be  
$B_{\rm tot} \equiv \sqrt{B_{\rm los}^2 + \left<B_{\rm pos}\right>^2 } \sim 154 \ \mu$G 
with an inclination angle of $\sim$45$^\circ$ with respect to the line-of-sight.
This value is comparable to the one derived from our Zeeman measurements.

Dynamical evolution of a magnetized core is determined 
by a dimensionless parameter called a normalized mass-to-magnetic-flux 
ratio \citep{crutcher12}.
\begin{equation}
\lambda \equiv {(M_{\rm core} / \Psi _B) \over 
(M_{\rm core} / \Psi _B)_{\rm cr}}  = {(N_{\rm H_2} / B) \over 
(N_{\rm H_2} / B)_{\rm cr}}  \  ,
\end{equation} 
where $N_{\rm H_2}$ is the column density of hydrogen molecules,
$M_{\rm core}$ is the core mass, $\Psi_B$ is the magnetic flux, 
%$G$ denotes the gravitational constant, and 
$(M_{\rm core} / \Psi _B)_{\rm cr}$ is the critical mass-to-flux ratio of 
$7.6\times 10^{-21} N_{\rm H_2}/B$ \citep{nakano78}.  
If $\lambda \ge 1$, the magnetic field is too weak to support the core against its contraction.
Such a core is called magnetically supercritical, and can continue to gravitational contraction
to form stars.  On the other hand, if $\lambda < 1$, the core is called magnetically subcritical.
In interstellar space, such a core can continue gravitational contraction by losing its magnetic 
flux due to ambipolar diffusion.  The timescale of the gravitational contraction 
is at least an order of magnitude longer than that of the former in the absence of
strong turbulence \citep{nakamura08}, 
and thus the resultant efficiency of star formation becomes very different from the former.  

From our derived field strength, the mass-to-flux ratio of TMC-1 is estimated
to be $\lambda \simeq 2.2$, 
where we used the H$_2$ column density of $3 \times 10^{22}$ cm$^{-2}$ 
from the {\it Herschel} data \citep{malinen12}.
Thus, TMC-1 is likely to have a moderately-strong supercritical magnetic field.
%plays an important role in dynamical evolution of the core.
Such a  magnetic field can remove the core angular momentum efficiently through the magnetic braking, but
cannot prevent from the gravitational contraction completely.
Our measurement is consistent with the results of the previous OH Zeeman observations \citep{troland08}
toward the position with about 5$'$ offset along the filament from our position 
(see figure \ref{fig:tmc-1}), corresponding to about 0.2 pc offset.
%There is another position where the magnetic field strength is measured on the basis
%of the OH Zeeman observations\cite{troland08}.
%The position has about 5$'$ offset along the filament from our position 
%(see figure 1), corresponding to about 0.2 pc offset. 
From the OH Zeeman observations, the field strength is derived to be $B_{\rm los} \simeq $ 14 $\mu$G at
 $n_{\rm H_2} = 5000$ cm$^{-3}$. 
If the field strength increases as $B_{\rm los} \propto n_{\rm H_2}^{2/3}$ (the weak field case),
the line-of-sight field strength is evaluated to be 46 $\mu$G, 
corresponding to $B_{\rm tot} \simeq 110 \ \mu$G when we adopt  $\left<B_{\rm pos}\right> \simeq $ 100 $\mu$G
with an inclination angle of $\sim$ 30$^\circ$.
%This field strength is significantly weaker than that of our observed position.
The flux-to-mass ratio at the OH position is estimated to be $\lambda \simeq 3$, 
where we adopt a column density of $2 \times 10^{22}$ cm$^{-2}$ \citep{malinen12}.
Thus, the OH position is magnetically supercritical, similarly to our position.

The infall velocity  at our position is inferred  to be  $\sim$ 0.6 km s$^{-1}$, about three times  the sound speed for $T=10$ K,
assuming that the infall proceeds along the magnetic field lines with an inclination angle of $\sim$ 45$^\circ$. 
This infall velocity is in agreement with the terminal velocity of isothermal collapse of spherical and cylindrical clouds \citep{larson69,nakamura98,ogino99}.
Thus, we suggest that the TMC-1 filament is now dynamically contracting toward the center and on the verge of protostellar formation.  
The dense cores distributed along the TMC-1 filament might be in the evolutionary stages just before the formation of the 
first hydrostatic core.

%The different magnetic strengths between our position and the OH position may infer 
%that the ambipolar diffusion reduced the flux-to-mass ratio in the eastern part of the TMC-1 filament, 
%whereas the eastern part is dynamically-younger and may just formed.
%According to the chemical evolution calculation and CCS/NH$_3$ observations\cite{suzuki92}, 
%TMC-1 may be as chemically-young as $\sim 10^4$ yr and 
%the western part of the TMC-1 filament is more chemically-evolved.
%%The column density at this position is somewhat larger than that 
%%of our observed position.
%If we assume one-to-one correspondence between chemical age and dynamical age,
%the different values of the flux-to-mass ratios at the two positions are consistent with an idea 
%that the dynamics of the TMC-1 filament is magnetically controlled and 
%the ambipolar diffusion reduced the magnetic flux at the western part. 
%TMC-1 may provide us a unique opportunity to investigate 
%how the ambipolar diffusion affects the dynamics of pre-stellar cores.

\section{Summary}
\label{sec:summary}

\begin{itemize}
\item[1.] 
We performed the CCS ($J_N=4_3-3_2$) Zeeman observations toward the TMC-1 using the Nobeyama 45-m telescope
and detected the significant splitting of 74.7 $\pm$ 13.0 Hz.
We estimated the \textcolor{black}{line-of-sight} magnetic strength to be 110 $\mu$G $\pm$ 21 $\mu$G.

\item[2.]
We also simultaneously observed a non-Zeeman line, HC$_3$N \textcolor{black}{to check how well the polarization calibration was perfomed}, and
we did not detect any significant shift in the Stokes $V$ profile. 
This non-detection of the HC$_3$N Zeeman shift
\textcolor{black}{supports} our detection of the CCS Zeeman shift.

\item[3.]
Using the H$_2$ column density measurements by {\it Herschel} observations, we estimated the normalized mass-to-magnetic flux ratio of 2.2, assuming the inclination angle of 45$^\circ$. The inclination angle is inferred from the comparison with the results of the near infrared linear polarization observations.
We conclude that the TMC-1 is magnetically supercritical.

\item[4.]
We recently suggested the TMC-1 infalling radially toward the filament axis by solving the detailed 
one-dimensional radiative transfer \citep{dobashi18}. 
This is consistent with our conclusion that the TMC-1 is magnetically supercritical.

\end{itemize}

\begin{ack}
This work is supported in part by a Grant-in-Aid for Scientific 
Research of Japan (24244017).
We thank Satoshi Yamamoto and Ryohei Kawabe for valuable comments, suggestions, 
and encouragements.
We are grateful to Hideo Ogawa,  Kimihiko Kimura, Yoshinori Yonekura, Shuro Takano, 
Daisuke Iono, Nario Kuno, Munetake Momose, Nozomi Okada, Minato Kozu, 
Yutaka Hasegawa, Kazuki Tokuda, Tetsu Ochiai, Taku Nakajima, and Hiroko Shinnaga 
for their contribution to developing the polarization system and valuable comments.
We are grateful to the staffs at the Nobeyama Radio Observatory
(NRO) for operating the 45-m. NRO is a branch of the National Astronomical 
Observatory, National Institutes of Natural Sciences, Japan.
\end{ack}

%\begin{methods}

\appendix

\section{Polarimetry}
Spectroscopic polarimetry is carried out using the dual linear polarization capability of the Z45 \citep{nakamura15} receiver and the PolariS spectrometer \citep{mizuno14}. PolariS is a polarization spectrometer that accepts four IF signals ($X_0$, $X_1$, $Y_0$, and $Y_1$; $X$ and $Y$ stand for linear polarization components and suffice for two spectral windows) with a 4-MHz bandwidth for each and produce four power spectra of $\left< X_0X^*_0 \right>$, $\left< X_1 X^*_1 \right>$, $\left< Y_0 Y^*_0 \right>$, and $\left< Y_1 Y^*_1 \right>$ and two cross-power spectra of $\left< X_0 Y^*_0 \right>$,  $\left< X_1 Y^*_1 \right>$. We assigned line species of CCS and HC$_3$N in the spectral windows of 0 and 1, respectively.
Stokes parameters are derived by using outputs from PolariS as
\begin{eqnarray}
\left(
\begin{array}{c}
\frac{\left< XX^* \right>}{G_X G^*_X}  \\
{\left< XY^* \right> \over G_X G^*_Y}  \\
{\left< X^*Y \right>  \over G_Y G^*_X}  \\
{\left< YY^* \right>  \over G_Y G^*_Y}  
\end{array}
\right)
= \frac{1}{2}
\left(
\begin{array}{cccc}
    1 & D^*_X & D_X & D_X D^*_X \\
D^*_Y &     1 & D_X D^*_Y & D_X \\
D_Y   & D_Y D^*_X &     1 & D^*_X \\
D_Y D^*_Y & D_Y & D^*_Y & 1
\end{array}
\right)
\left(
\begin{array}{cccc}
1 & \cos 2\psi & \sin 2\psi & 0 \\
0 &-\sin 2\psi & \cos 2\psi & i \\
0 &-\sin 2\psi & \cos 2\psi & -i \\
1 &-\cos 2\psi &-\sin 2\psi & 0 
\end{array}
\right)
\left(
\begin{array}{c}
I  \\
Q  \\
U  \\
V  
\end{array}
\right). \label{eqn:StokesParameters}
\end{eqnarray}
Here $\psi$ is the position angle of the $X$-polarization feed projected on the celestial sphere. With the Nasmyth optics of the Nobeyama 45-m telescope, it is given by $\psi = \phi_{\rm PA} + \phi_{\rm EL}$, where $\phi_{\rm PA}$ and 
$\phi_{\rm EL}$ are parallactic angle and elevation angle, respectively.
The first matrix in equation (\ref{eqn:StokesParameters}) is composed by D-terms, which indicates cross talk coefficients between $X$ and $Y$.
The voltage-based complex gain is described as $G_X$ and $G_Y$. The argument of $G_Y$ indicates relative phase between $X$ and $Y$ polarizations.

We address Stokes $V$ which responses the Zeeman split.
When the source is not linearly polarized (i.e. $Q = U = 0$), we have
\begin{eqnarray}
{\left< XY^* \right> \over G_X G^*_Y} = (D_X + D^*_Y)I + i(1 - D_X D^*_Y) V. \label{eqn:XYcorrelation}
\end{eqnarray}
%Calibration to measure Stokes $V$ is carried out by determining complex gains and D-terms.
The instrumental phase offset between $X$ and $Y$ polarizations, measured using the wire grid 
and the P-cal signal (see Section 2), is stored in $G_Y$. This $X$-$Y$ phase calibration is verified by non-detection of Stokes $V$ toward the Crab nebula.

\section{Gain and phase calibration}
The amplitudes of gains can be determined by applying system noise temperature, $T_{\rm sys}$ 
as 
\begin{equation}
|G| = {1\over \sqrt{2 k_B T_{\rm sys}}}  \ ,
\end{equation}
where $k_B$ is the Boltzman constant.
We used the digital measuring method \citep{2010PASJ...62.1361N} using level histograms of the digitizer in the PolariS.
Because Stokes $V$ component is given by the imaginary part of $\left< XY^* \right>$, it is crucial to calibrate the phase of $G_Y$, with respect to that of $G_X$. We employed three ways of $G_Y$ phase calibration; a linearly polarized calibrator source, wire grid, and phase-calibration (P-cal) signal.
We used the Crab nebula as the linearly polarized calibrator source whose polarization angle is $151^{\circ}.0$ at 31 GHz \citep{2005ApJ...623...11C} or $152^{\circ}.1 \pm 0^{\circ}.3$ at 90 GHz. To determine the orientation of $X$ and $Y$ feed projected on the sky, we observed the Crab nebula before and
after the meridian at the different parallactic angles.
In addition, we inserted a wire grid in front of the receiver feed to input linearly polarized signals. While signals from the sky transmit through the wire grid, the reflected signals originate from the ambient load (i.e. an absorber at the room temperature). The contrast between the temperatures of the ambient load and the sky yields linearly polarized continuum signals.
Cross correlation of $\left< XY^* \right>$ while inserting the wire grid allows us to calibrate $XY$ delay and phase.
The wire grid was inserted during antenna slew before calibrator scans.
To verify time stability of $XY$ phase, we also injected a phase calibration signal (P-cal) which is a linearly polarized monochromatic wave at the frequency beside the CCS emission.
We solve 180-degree ambiguity in phase of P-cal by comparing with cross correlation toward the Crab nebula. 
%Because we have 180-degree ambiguity in the phase of P-cal, its phase is compared with the cross %correlation toward the Crab nebula.

We measured the delay and phase offsets between $X$ and $Y$ signal paths by taking cross correlation, $\left< XY^* \right>$, when we inserted a wire grid in front of the feed horn (see figure 12 in \citet{nakamura15}). The reflection and transmission waves at the wire grid terminate 
onto the ambient load and the sky, respectively. Thus, the amplitude of the normalized cross correlation coefficient is 
$(T_{\rm amb}- T_{\rm sky})/\sqrt{(T_{\rm RX} + T_{\rm sky})(T_{\rm RX} + T_{\rm amb})}$, and the phase, 
$\phi = \phi_{XY} + 2 \pi \tau_{XY} \nu_{BB}$, where $\phi_{XY}$ and $\tau_{XY}$ are phase and delay offsets between $X$ and $Y$ signal paths, respectively.

\section{D-term correction}
The cross polarization leakage (D-term) is determined by using both unpolarized calibrators and linearly polarized calibrator, respectively.
We used the Venus or Jupiter as the unpolarized calibrator. We assume that Stokes $Q=U=V=0$ for the unpolarized calibrator, then the cross correlation will be $\left< XY^* \right> / G_X G^*_Y = (D_X + D^*_Y)I$. Giving the model Stokes $I$ (antenna temperature) of the calibrator, we determine the value of $D_X + D^*_Y$ and apply it to equation (\ref{eqn:XYcorrelation}).

\section{Beam squint}
Although we employed dual linear polarization feeds, we focus on Stokes $V$ which is proportional to the difference between circular polarizations (RHCP and LHCP). The optics of the antenna and the receiving system may have differences in beam patterns between RHCP and LHCP.
The inhomogeneity in circular polarization can cause false Stokes $V$, and then produce fake Zeeman-split-like features when the source has a velocity gradient.
The fake Zeeman shift, $\Delta v$, is given by
\begin{eqnarray}
\Delta v = 
\left(
\begin{array}{c}
\frac{dv}{d\alpha}  \\
\frac{dv}{d\delta}
\end{array}
\right) 
\left(
\begin{array}{cc}
\cos {\rm \phi_{\rm PA}} & \sin {\rm \phi_{\rm PA}}  \\
-\sin {\rm \phi_{\rm PA}} & \cos {\rm \phi_{\rm PA}}
\end{array}
\right)
\left(
\begin{array}{c}
\theta_{\rm AZ}  \\
\theta_{\rm EL} 
\end{array}
\right), \label{eqn:velocityGradient}
\end{eqnarray}
where $\frac{dv}{d\alpha}$ and $\frac{dv}{d\delta}$ are the velocity gradients, $\phi_{\rm PA}$ is the parallactic angle, and $(\theta_{\rm AZ}, \theta_{\rm EL})$ is the beam squint between RHCP and LHCP.
We measured the beam squint by observing CH$_3$OH maser (44.1 GHz) emission toward the OMC-2 \citep{sarma11, 2012AJ....144..189M}.
The maser source is so compact that we presume the identical positions of RHCP and LHCP emissions.
Cross scans towards the maser yields the beam squint of $(\theta_{\rm AZ}, \theta_{\rm EL}) \simeq (-2^{\prime \prime}.08 \pm 0^{\prime \prime}.30, -0^{\prime \prime}.24\pm 0^{\prime \prime}.30)$.

\begin{figure}
\includegraphics[width=120mm, bb = 0 0 600 600]{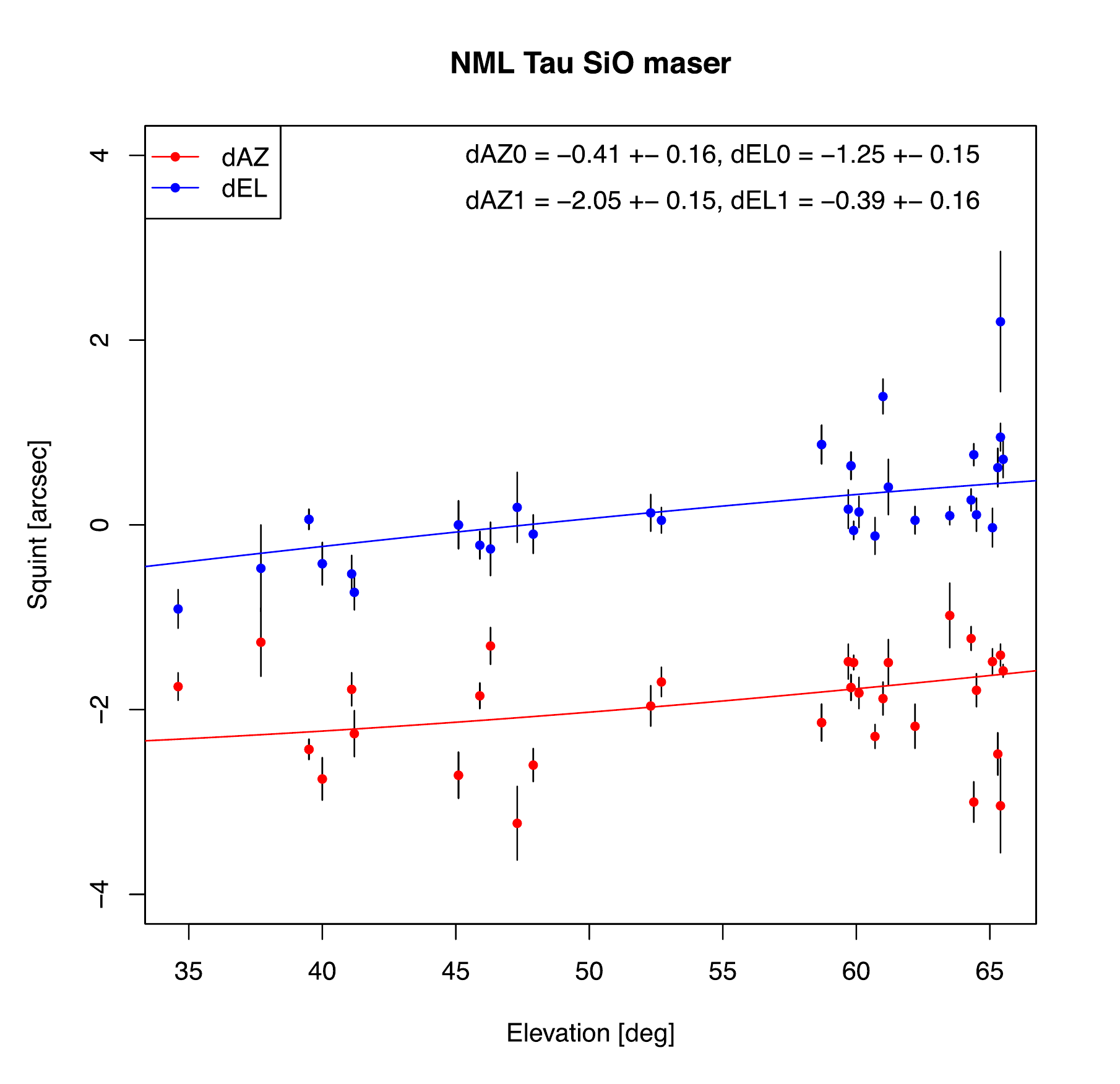}
\caption{The beam squint measured with the SiO maser line from NML Tau as a function of the observed elevation.  
The fitted functions are shown in the upper right.}
\label{fig:beam}

\end{figure}

\begin{figure}
\includegraphics[width=120mm, bb = 0 0 600 600]{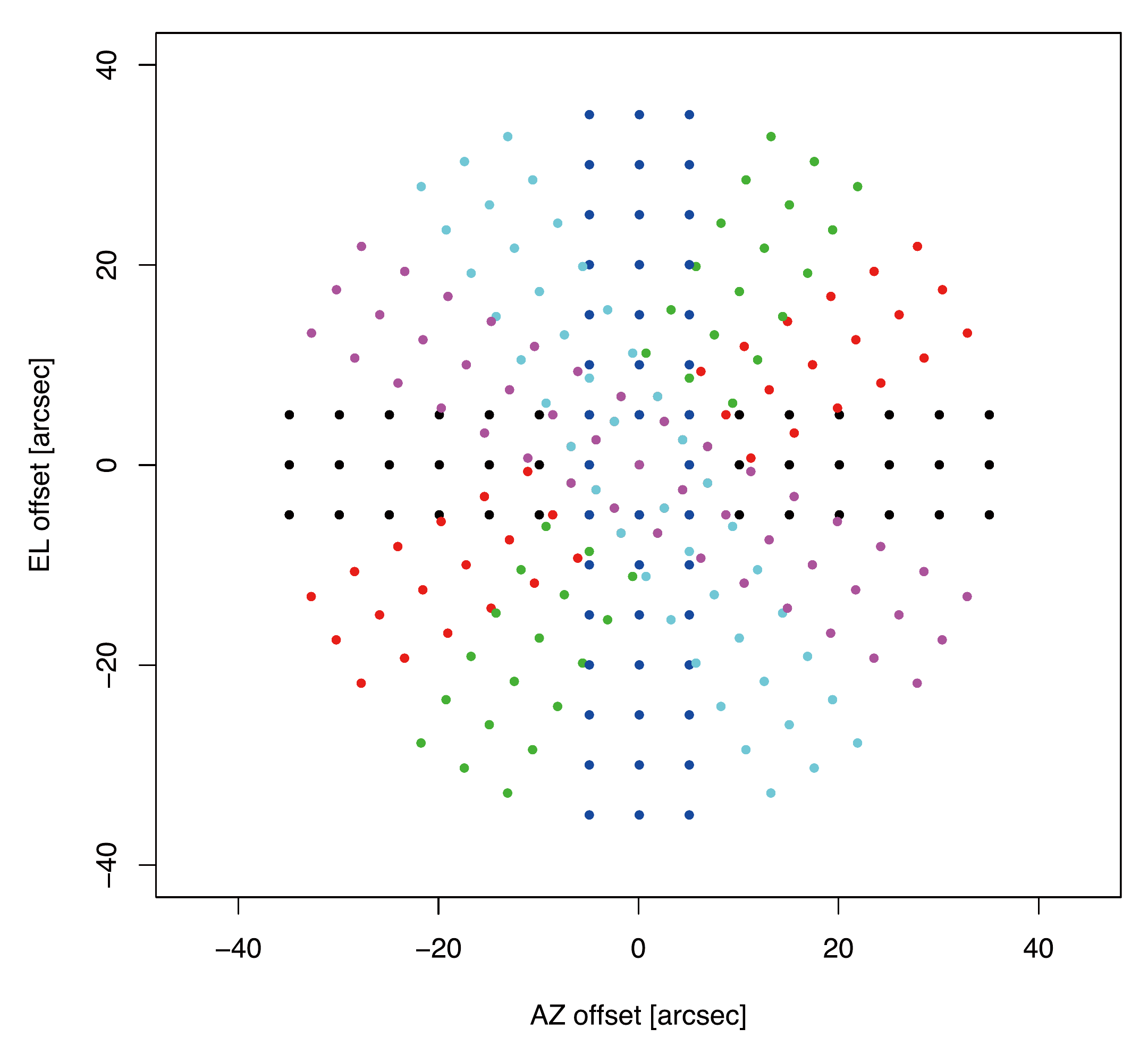}
\caption{\textcolor{black}{Scan pattern of the beam-squint measurement toward the NML Tau SiO maser. Rectangular 15 $ \times$ 3 grid spaced by 5\arcsec (i.e., 70\arcsec $\times$ 10\arcsec) excursions are rotated by 30$^\circ$ step to cover  a 35\arcsec-radius circle centering the SiO maser.}}
\label{fig:scan}
\end{figure}

More accurately, the beam squint has a dependence on the elevation of the observations. 
Figure \ref{fig:beam} indicates the beam squint measured with the SiO maser line toward NML Tau 
as a function of the observed elevation.  We fit the beam squint of ($dAZ, dEL$) as
\begin{eqnarray}
dAZ &=& dAZ_0 + dAZ_1 \cos EL + dEL_1 \sin EL  \\
dEL &=& dEL_0 - dAZ_1 \sin EL + dEL_1 \cos EL  
\end{eqnarray}
where $dAZ_0 = -0^{\prime \prime}.41 \pm 0^{\prime \prime}.16$,
$dEL_0 = -1^{\prime \prime}.25 \pm 0^{\prime \prime}.16$,
$dAZ_1 = -2^{\prime \prime}.05 \pm 0^{\prime \prime}.15 $, and
$dEL_1 = -0^{\prime \prime}.39 \pm 0^{\prime \prime}.16$.
towards rectangular mapping towards the SiO maser source NML Tau.
\textcolor{black}{The scan pattern is presented in figure \ref{fig:scan}.}
We applied correction in the calibration process of the beam squint using the above fitting formula.

We also measured the velocity gradient of CCS and HC$_3$N emission by analyzing mean velocity (moment-1)  map of the TMC-1 molecular cloud. We cropped the field where the signal-to-noise ratio of the integrated intensity (moment-0 map) exceeds $2\sigma$ of the standard deviation of emission-free area. The velocity gradients of CCS and HC$_3$N were given as $(3.68, 3.43)$ km s$^{-1}$ deg$^{-1}$ and $(3.63, 2.74)$ km s$^{-1}$ deg$^{-1}$, respectively.
Then every Stokes $V$ spectrum was corrected by applying the beam squint and the velocity gradient.

\section{\textcolor{black}{Four components of CCS}}

The CCS line profile at our observed position appears to have multiple components
with different line-of-sight velocities [see Dobashi et al.(2018) for more accurate and detail analysis].  
The CCS emission is relatively strong and 
therefore emission from components in the back can be absorbed by molecules
in the foreground components.
Applying radiative transfer calculations, we can infer the spatial configuration
of the multiple components along line-of-sight.

We observed CCS ($J_N=4_3-3_2$) and HC$_3$N ($J=5-4$) lines simultaneously.
The HC$_3$N ($J=5-4$) line has multiple hyperfine structure.
Two hyperfine lines of $F=5-5$ and $F=4-4$ are weak and are isolated from other 
hyperfine lines ($F=4-3$, $F=5-4$, $F=6-5$). Thus, intrinsic line intensities of these hyperfine lines are likely to be optically thin.
Since HC$_3$N ($J=5-4$) has a critical density comparable to that of CCS ($J_N=4_3-3_2$), 
we consider that they trace molecular gas with a similar density range.

First, we fit the HC$_3$N ($J=5-4$, $F=5-5$ and $F=4-4$) with $n$
Gaussian components. Each component has three free parameters: the peak intensity,
centroid velocity, and velocity dispersion. In other words, we have $3n$ free parameters.
We fit the line profile in the range of $V_{\rm LSR}=5.2-6.2$ km s$^{-1}$ 
applying the $\chi ^2$ test by increasing the number of $n$.
The  reduced $\chi^2$ of the fit obtained $\chi_r^2$=2.6543, 1.0896, 1.0016, and 1.0013 for $N=2,3$, 4, and 5, respectively. 
The $\chi^2_r$ value becomes almost $\chi^2_r \simeq 1$ for $N\ge 4$. In addition, for $N\ge 5$, the Gaussian components tend to fit
the noises in the spectral line profile. 
Thus, we found that the line profile is fitted with $N=4$ Gaussian components, 
and we call these components A--D in the order of the best fitting centroid velocities. 
We show the results of the fitting in figure \ref{fig:fitted}(a).

%\begin{table}
%  \title{The fitted parameters for the HC$_3$N ($J=5-4$, $F=5-5$ and $F=4-4$) Line }{%
%  \begin{tabular}{llll}
%  \hline
%  Component & Peak Intensity & The Centroid Velocity & Velocity Dispersion  \\
%   & (K) & km s$^{-1}$ & km s$^{-1}$  \\
%\hline
%A    & 0.48 $\pm$ 0.02 & 5.727$\pm$ 0.002 & 0.054$\pm$ 0.001 \\
%B & 0.60 $\pm$ 0.02 & 5.901$\pm$ 0.007 & 0.088 $\pm$0.009\\
%C & 0.35 $\pm$ 0.08 & 6.064$\pm$ 0.007 & 0.061 $\pm$0.012\\
%D & 0.31 $\pm$ 0.07 & 6.160$\pm$ 0.006 & 0.047 $\pm$0.002\\
%\hline
%  \end{tabular}}
%   \label{tab:gaussian}
%% \begin{tablenote}
%%The peak intensity is given in the antenna temperature scale.
% %  \end{tablenote}
%\end{table}

Next, we fit the CCS line profile with 4 components with the same
centroid velocities as we obtained from HC$_3$N.
We assume that the optical depth of each component has a Gaussian shape.
\begin{equation}
\tau _i (v) = \tau ^0 _i \exp \left[-{1\over 2} \left({v-v_i \over \sigma_i} \right) ^2\right]  \  ,
\end{equation}
where $\tau_i^0$, $v_i$, and $\sigma_i$ are the optical depth at the line center,  centroid velocity, and
the velocity dispersion of the $i$-th component, respectively.
The intrinsic emission of the $i$-th component, $T_i$, before being absorbed by the foreground gas is expressed as
%The emission of the $i$-th component, $T_i$, is computed as
\begin{equation}
T_i (v)=T_i^0 [1-\exp(-\tau _i (v))]  \  ,
\end{equation}
where $T_i^0$ is the intensity at the line center for $\tau _i^0 \gg 1$. If we assume that all of the four components are 
lying on the same line-of-sight in the order of  $i=1, 2, 3$, and 4 having the $i=1$ component to be the furthest 
from the observer, the total emission we should observe is then given as
\begin{equation}
T =\sum _{i=1} ^{4} T_i [1-\exp (- S_i (v))]  \ ,
\end{equation}
where
\begin{equation}
S_i (v) =
\left\{ \begin{array}{ll} \sum_{j=i+1}^4 \tau_j(v) & (i = 1, 2, 3) \\ 0 & (i=4) \end{array} \right.   \ .
%\sum _{j= i+1} ^4 \tau_j (v) \hspace{5mm}   (i=1,2,3)  \\
%&=& 0 \hspace{2cm} (i=4)
\end{equation}
We fit the CCS line profile with the above computed  total emission, $T$.
In total, we have $4!=24$ possible spatial configurations 
of the four components A--D to assign $i=1$ (the furthest) to $i=4$ (the nearest, see figure \ref{fig:components}). 
Each configuration has $3 \times 4 =12$ free parameters 
($\tau _i ^0$ and $\sigma _i$ in Equation (A7) and $T_i^0$ in Equation (A8)).
%of four components.
%Each configuration has free parameters of 8 (the optical depth at the line center
%and velocity dispersion of each component).  
Applying the $\chi ^2$ test, we found the most plausible spatial configuration is 
(A, B, C, and D) from the furthest position along the line-of-sight giving a reduced $\chi^2$ of $\sim 1.06$. 
The results of the $\chi^2$ test are summarized in Table \ref{tab:chi2}.
We show the fitted spectrum in figure \ref{fig:fitted}(b).
We also summarized the fitting parameters in Table \ref{tab:gaussian}.
Though the true configuration of the four velocity components may not be as simple as in our model, 
we would interpret the results of the radiative transfer calculations 
as the global motion of contraction.

\citet{dobashi18} conducted essentially the same calculations as above including 
the HC$_3$N ($J=4-3$, $F=5-4$ and $F=6-5$) main components which are optically thicker than the CCS line.
Therefore, the spatial configuration is more distinguishable due to stronger absorption effects.
Their result agrees well with that of CCS.  
Thus, we concluded that the most plausible spatial configuration is (A, B, C, and D) from the furthest position along the line-of-sight.

%Because the D and A components do not have significant overlap in spectra, 
%it is difficult to constrain the positions of D and A from the radiative transfer calculations.
%However, we interpret that these configurations may suggest the existence of global motion of
%either expansion or contraction.

\section{\textcolor{black}{Stokes $I$ and $V$ spectra for the main component of HC$_3$N}}

\begin{figure}
\includegraphics[width=80mm, bb = 0 0 800 800]{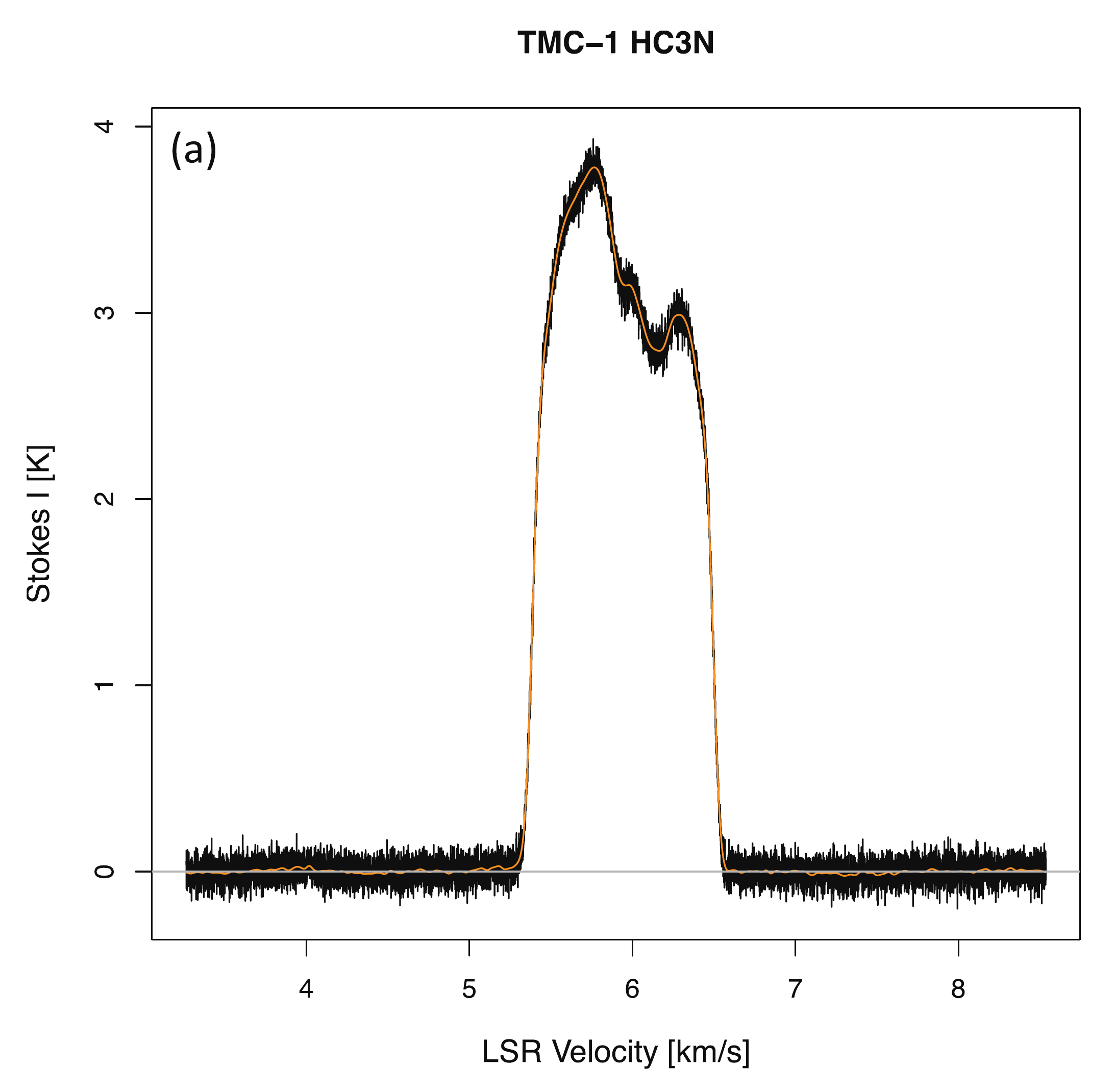}
\includegraphics[width=80mm, bb = 0 0 800 800]{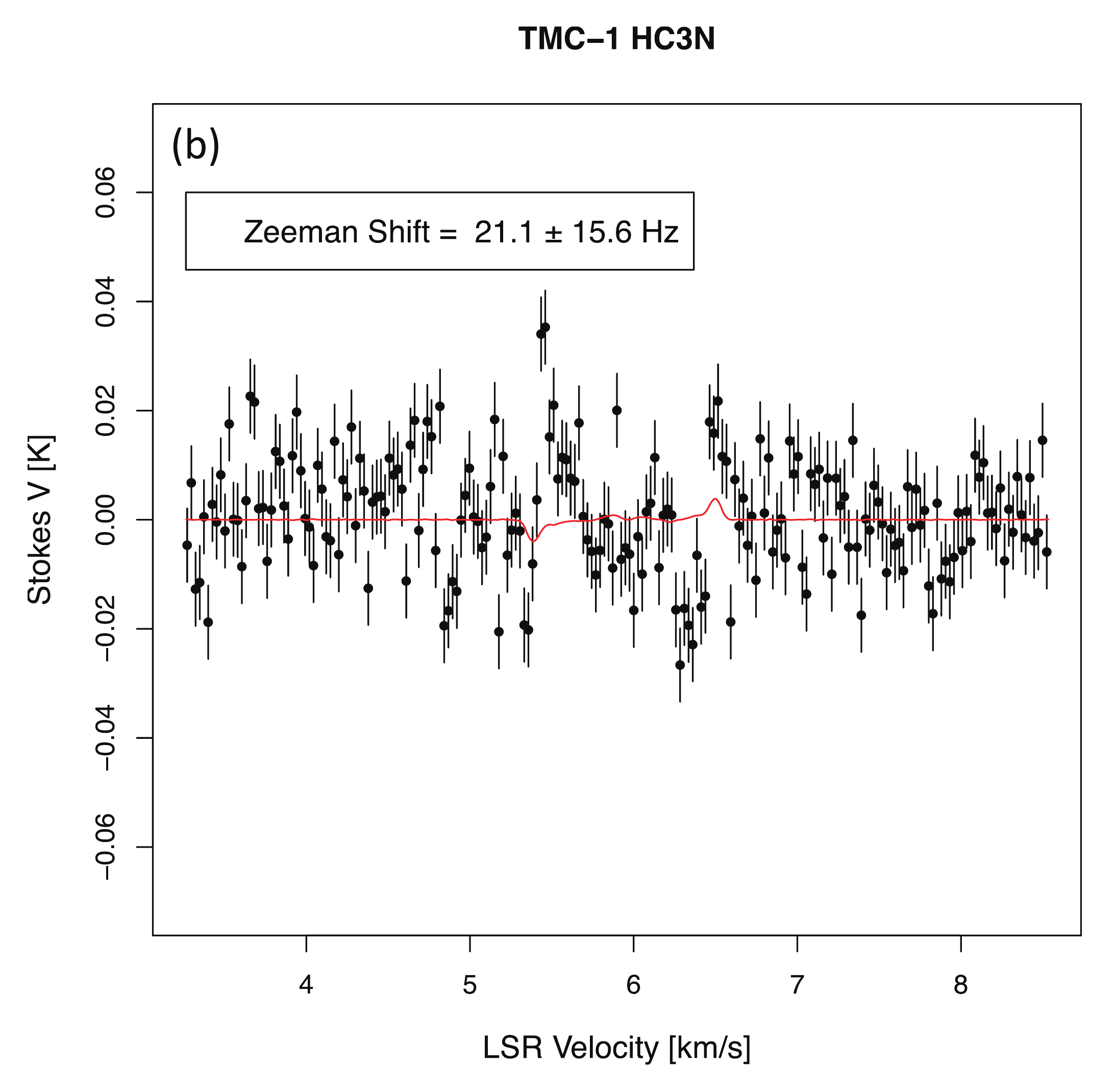}
\caption{\textcolor{black}{Same as figure  \ref{fig:hc3n} but for the HC$_3$N  ($J=5-4$) main component 
($F=6-5, 5-4, 4-3$) toward TMC-1.
The three hyperfine components are blended for the main component, and thus 
the main line consists of 12 components since 4 components with different velocities are blended 
at the observed position.
(b) The Stokes $V$ profile of the main component of HC$_3$N. The red line shows the fitted Stokes V profile.
The $t$ and $p$ values of the fitting were $t=0.7$ and $p=0.48$, respectively.}}
\label{fig:hc3n_main}
\end{figure}

\textcolor{black}{For confirmation, we applied the same procedure as used for the CCS analysis to 
the main component of HC$_3$N. The Stokes $I$
and $V$ profiles are shown in figure \ref{fig:hc3n_main}. The main component of HC$_3$N 
consists of three hyperfine components. The TMC-1 position contains 4 components with 
different velocities and thus 12 components are blended for HC$_3$N.  The 4 components may have different optical depths and different line-of-sight velocity gradients.
These effects are likely to cause the different beam squints, and presumably sometimes significantly influence the derived Stokes $V$ pattern when the velocity gradients of blended components are very different. }

\textcolor{black}{
Therefore, for the verification of our polarization calibration, we used a satellite line, which does not contain different hyperfine components. 
We note that the $\chi^2$ test for the fitting of the main components shows no 
statistically-significant shift, and thus 
we can conclude that our polarization calibration is reasonably accurate  even for the main components.
However, we recommend to use relatively-strong non-Zeeman emission lines, which do not contain blended hyperfine lines, for the similar verification of the polarization calibration. 
For the measurements toward other positions, this effect was likely to generate artificial 
patterns in the derived Stokes $V$ profiles for the main component.}

\clearpage

\begin{table}
\footnotesize
\caption{The reduced $\chi^2$ value for the different configurations}
  \title{The reduced $\chi^2$ value for the different configurations}{%
  \begin{tabular}{ll}
  \hline
Configuration & $\chi^2_r$  \\
   &   \\
\hline
ABCD   & 1.060 \\
ABDC & 1.114\\
ACBD & 1.124\\
ACDB &1.147\\
ADBC & 1.104\\
ADCB & 1.081\\
BACD &1.085\\
BADC & 1.157\\
BCAD& 1.085\\
BCDA& 1.085 \\
BDAC& 1.157 \\
BDCA& 1.157\\
CABD & 1.127\\
CADB &1.147\\
CBAD&1.097\\
CBDA& 1.097\\
CDAB& 1.147\\
CDBA& 1.102\\
DABC &1.105\\
DACB &1.081\\
DBAC &1.161\\
DBCA &1.161\\
DCAB &1.081\\
DCBA &1.102\\
\hline
  \end{tabular}}
   \label{tab:chi2}
% \begin{tabnote}
%The peak intensity is given in the antenna temperature scale.
 %  \end{tabnote}
\end{table}

\begin{table}
\caption{The fitted parameters for the CCS ($J_N=4_3-3_2$) Line}
  \title{The fitted parameters for the CCS ($J_N=4_3-3_2$) Line }{%
  \begin{tabular}{lllll}
 \hline
  Component & Peak Intensity ($T_i^0$) & Velocity Dispersion ($\sigma_i$) & 
  $\tau_i^0$ & $T_{\rm ex}$  \\
   & (K) & (km s$^{-1}$) &  & (K) \\
 \hline
A    & 2.22 $\pm$ 0.04 & 0.052 $\pm$ 0.001 & 2.15 $\pm$0.07 & 6.0 $\pm$0.1\\
B & 3.17 $\pm$ 0.07  & 0.116$\pm$ 0.001 & 1.37 $\pm$ 0.07 & 7.3$\pm$ 0.2\\
C & 2.20 $\pm$ 0.01  & 0.070 $\pm$0.002& 1.88 $\pm$ 0.10 & 5.9$\pm$ 0.1\\
D & 2.79 $\pm$ 0.03  & 0.047 $\pm$0.001 & 1.08 $\pm$ 0.05 & 6.8$\pm$ 0.1\\
\hline
  \end{tabular}}
   \label{tab:gaussian}
% \begin{tabnote}
%The peak intensity is given in the antenna temperature scale.
 %  \end{tabnote}
\end{table}

%
%\subsection{Stokes I and I spectra for the main component of HC$_3$N}
%
%For confirmation, we applied the same procedure as used for CCS analysis to 
%the main component of HC$_3$N and HC$_5$N.
%The Stokes I and V profiles are shown in figure \ref{fig:hc3n_main}. 
%The main component of HC$_3$N consists
%of three hyperfine components.  The TMC-1 position contains 4  components with different
%velocities and thus 12 components are blended for HC$_3$N.
%The $\chi^2$ test of the fitting shows no statistically significant shift for the two lines.
%Note that we observed HC$_5$N and CCS simultaneously. The total time of these observations  
%are shorter than the HC$_3$N and CCS set, and thus signal-to-noise ratio is not so good as 
%the HC$_3$N/CCS observations.

\clearpage

\end{document}